# Correlations and strong interactions[*]

R.M.Weiner [§]

Physics Department, University of Marburg, Marburg, Germany

Although Quantum Chromodynamics (QCD) is considered at present to be the best candidate for a theory of strong interactions, it is not directly applicable to the most interesting and important aspects of strong interactions, i.e. to multiparticle production, which is a "soft" process. Instead one uses lattice QCD which predicts in principle the hadronic mass spectrum. The fact that the main content of Hagedorn's statistical bootstrap is also the mass spectrum proves the intuitive power of this approach, developed many years before the advent of QCD.

Why is the mass spectrum so essential for the understanding of strong interactions? In the Hagedorn theory the answer lies in the very derivation of this spectrum. The bootstrap "closes" only for a specific analytical form of this spectrum (the exponential one) and the fact that the bootstrap closes is the reflection of the strong nature of the interaction.

In the following I shall provide a further argument for the importance of the mass spectrum by relating it to another important phenomenological characteristic of strong interactions, namely the multiplicity distribution $P(n)$. I shall then show that the knowledge of this distribution is equivalent with the knowledge of correlations. The remainder of the talk will be devoted to a short review of the most important developments in the field of correlations in the last 5 years.

## 1 Multiplicity distributions and correlations

In field theory various interactions are characterized by the values of the corresponding coupling constants. Phenomenologically the coupling constant manifests itself in the magnitude of cross sections and in the multiplicities and their distributions.

---

[*]Invited talk at the Workshop Hot Hadronic Matter, Divonne, June 1994

[§]E.-mail address: weiner@vax.hrz.uni-marburg.de

Thus at a given energy of the system the mean multiplicity in electroweak interactions (ew) is much smaller than the corresponding one in strong interactions (s) $\langle n \rangle_{ew} \ll \langle n \rangle_s$. For the multiplicity distributions the corresponding inequality reads $P_{ew}(n) \ll P_s(n)$ (for $n > 1$). Furthermore we have $P_{ew}(n+1) \ll P_{ew}(n)$ to be compared with $P_s(n+1) \simeq P_s(n)$. In words the last relation[1] states that in strong interactions the probability to produce $n+1$ particles is comparable with the probability to produce $n$ particles and this can be considered as a phenomenological definition of strong interactions.

Rather than working with multiplicity distributions it is often convenient to consider their moments, to which they are mathematically equivalent. In particular let us consider the second order normalized factorial moment $f_2 = \langle n(n-1) \rangle / \langle n \rangle^2$. Here the averaging is performed with respect to the distribution $P(n)$ i.e. $\langle x^q \rangle = \sum_{x=1}^{\infty} x^q P(x)$. This moment is related to the second order correlation function

$$C_2(k_1, k_2) = \rho_2(k_1, k_2) / [\rho_1(k_1) \rho_1(k_2)] \tag{1}$$

by the equation

$$f_2 = \frac{\int C_2 \rho_1(k_1) \rho_1(k_2) d^3k_1 d^3k_2}{\int \rho_1(k_1) \rho_1(k_2) d^3k_1 d^3k_2} \tag{2}$$

In Eq.(2) $\rho_1$ and $\rho_2$ are the single and double inclusive cross sections. Similar relations hold for the higher order moments and correlations. Herefrom follows that *multiplicity distributions are determined by the correlations* and from the arguments presented above we conclude that correlations, too, reflect the nature of interactions. As a matter of fact this conclusion could have been reached without going through the multiplicity distributions because lattice calculations provide directly correlation lengths from which the masses of particles are derived.[2] We used the small detour through multiplicity distributions because multiplicity distributions and their moments are important physical quantities in themselves which have been in the center of interest of soft strong interaction phenomenology. They demand different experimental measurements than correlations or mass measurements and are treated usually separately and quite often the link between these two categories is overlooked.

---

[1] Here we assume that the energy is high enough so that rest masses do not matter.

[2] So far the correlations obtained from lattice calculations refer only to quark-antiquark states, but in principle more complex correlations between pairs of $q - \bar{q}$ states i.e. meson-meson correlations should be obtainable.

# 2  New developments in the field of correlations

## 2.1  Long range versus short range correlations

It is customary to distinguish between long range correlations (LRC) and short range correlations (SRC) in momentum (rapidity) space. Although this distinction is not clear cut, in a first approximation one considers $Q = 1 GeV$ as the dividing line between these two types of correlations. Here $Q$ is the invariant four momentum difference between two particles. In the eighties the change with energy of (normalized) $P(n)$ was discovered (KNO scaling violation) and different phenomenological models for this effect were proposed (cf.e.g.[1]), but in the following I shall mention only the quantum optical approach because it applies both to SRC and LRC. In the quantum optical approach one parametrizes the correlation in terms of a coherence length $\xi$ and chaoticity $p$ by writing the correlator of the (pionic) fields $\pi$

$$\langle \pi(y_1)\pi(y_2)\rangle = \langle n_{ch}\rangle exp(-|y_2 - y_1|/\xi) \qquad (3)$$

$\langle n_{ch}\rangle$ is the mean number of chaotic particles related to the chaoticity $p$ and the mean total multiplicity $\langle n \rangle$ by the equation $p = \langle n_{ch}\rangle/n$. The coherence length $\xi$ which refers to fields is related to the conventional correlation length $\lambda$ which refers to intensities (i.e. numbers of particles) by the equation $\lambda = \xi/2$. The broadening of the multiplicity distribution with energy was interpreted in this quantum optical approach as an increase of the correlation length and chaoticity with energy[2]. At that time only the UA5 data [3] for multiplicity distributions of charged particles were available and it could not be tested whether this interpretation applies indeed also to identical particles where Bose Einstein correlations (BEC) come into the game and in which SRC play a major part. With the advent of the newly analyzed UA1 [4] and Na22 [5] data this situation has changed. It turns out that the asymptotitc values of the correlations functions for large Q coincide with the values of the respective moments, which according to eq.(2) proves that: 1.the small Q region, i.e. SRC do not contribute to the moments; 2. *the increase of moments with energy is due to the increase of* LRC [6].

Property 2. can be due either to a change of the correlation length $\xi$ as suggested previously or to a change of the inelasticity (impact paramenter) or number of sources distribution or a combination of these. It appears that the last possibility is realized in nature and that the intensity and range of LRC increase with energy.

It is remarkable that conclusion 2. was reached by studying simultaneously multiplicity distributions *and* BEC which were considered up to that moment as refering almost exclusively to SRC. Further developments in the field of SRC and in particular in BEC will be discussed in the following.

The natural question to ask after this finding was whether this exhausted the energy dependence of correlations or whether there is also an energy dependence of SRC . It turns out that SRC and in particular BEC do also depend on energy.

## 2.2 Bose Einstein correlations

### 2.2.1 Correlation length versus radius

Everybody knows that BEC can be used for the determination of radii and lifetimes of sources. What is less known and has been clarified only recently[7] is the fact that actually in BEC at least two distinct length scales enter: a "radius" R and a correlation length L.The length L characterizes the region over which the generating currents $J$, responsable for the chaotic fields, are correlated. It defines thus in some sense the range of SRC. It is given by the space-time correlator

$$\langle J(x_1)J(x_2)\rangle \equiv C(x-y) = C(0)\exp[-\frac{(x_0-y_0)^2}{2L_0^2} - \frac{(\vec{x}-\vec{y})^2}{2L^2}] \tag{4}$$

for a "static", i.e. velocity independent source and by

$$C \propto exp[-(2(\tau_0^2/L_\eta^2)\sinh^2[(\eta_1-\eta_2)/2]] \tag{5}$$

for an expanding source. Here $\tau_0$ is the proper formation time, $\eta$ is the space- time rapidity and $L_\eta$ the corresponding correlation length. For simplicity I have not written out explicitly in the last equation the transverse coordinate dependence and have assumed $\tau_1 = \tau_2 = \tau_0$.

To obtain the correlation function one has still to define the space-time distribution of the source which is characterized by the four-radius $R$. A typical form for such a distribution reads

$$f(x) \propto exp[-x^2/R^2] \tag{6}$$

With the definitions

$$F(k_1,k_2) = \int \frac{d^4k}{(2\pi)^4} f^*(k_1+k)C(k)f(k+k_2), \quad k_r^0 = E_r, \tag{7}$$

$$d_{12} = \frac{F(k_1,k_2)}{[F(k_1,k_1)\cdot F(k_2,k_2)]^{\frac{1}{2}}} \tag{8}$$

where we used the same symbols for the Fourier transforms of $f$ and $C$, the second order correlation function reads

$$C_2^{++}(k_1, k_2) = 1 + |d_{12}|^2 \qquad (9)$$

The two particle inclusive cross section depends now both on the radius $R$ and on the correlation length $L$. Applying this formalism to the data at $\sqrt{s} = 22 GeV$ and $630 GeV$ one finds [8] that *R increases* with the energy while *L decreases*. It appears therefore that *the range of SRC decreases with s while the range of LRC increases.*

Subsequent to the introduction of the correlation length described above, Sinyukov (cf. ref.[9]) introduced a similar concept in statistical hydrodynamics. He uses the name "length of homogeneity" for it and it is easy to see that it plays in his hydrodynamical approach to BEC the same role as does the correlation length in the classical current approach to BEC.

### 2.2.2 Intermittency and BEC

This topic has preoccupied the high energy physics community for some time. It started with the suggestion by Bialas and Peschanski [10] that the dependence of factorial moments on the width of the rapidity distribution might reflect an intermittency property and therefore be power like. The fact that experimental data showed such a behaviour was interpreted by some authors as evidence for intermittency. Quite soon, however, it was realized that Bose Einstein correlations would lead to similar effects[11]. This possibility has been definitely confirmed by newer data where it was found that only identical particles showed the "intermittency" property. This is true both for the moments [12] as well as for the correlations of various order.

In order to cope with this observation, Bialas [13] introduced the new idea that the sources have fractal structure and therefore BEC which measure radii of sources show power dependence as a function of, say, $Q$. It is worth mentioning that the powerlike radius distribution introduced in [13] contains a cut-off length the origin of which is unknown. This cut-off length breaks the scaling behaviour and the comparison with data indicates that it corresponds to a momentum difference of 30-40 MeV, i.e. it represents a length of the order of 6-7 fm. For a hadronic system this is an exceedingly large value (cf. also [7]) and this fact alone may cast doubts about the validity of the entire idea. Nevertheless the concept of a fractal source is a very attractive one and to test this intriguing possibility, new analytical and numerical calculations were performed [14]. We started from a conventional source with a fixed size (eq.6) and calculated the corresponding correlation function at given Q. The resulting $C_2(Q)$ showed as a function of Q power behaviour. The analytical study of $C_2(Q)$ suggests that this happens as a consequence of the cylindrical phase space. Furthermore good agreement with the experimental UA1 data, which were the basis of the considerations of [13], is obtained. Last but not least all parameters which enter this calculation are well understood physical quantities (e.g. the typical radii and correlation lengths involved are all of the order of 1-fm) and no exotic cut-off is necessary.

We conclude herefrom that conventional BEC with sources of *fixed* i.e. non fractal size, can account for the present experimental observations and that there is no need for intermittency assumptions. This may be a depressing conclusion for many experimentalists and theorists (the number of papers on this subject is a 3 digit figure!) but appears almost unavoidable.

### 2.2.3 Resonances and Bose-Einstein correlations

The majority of secondaries produced in high-energy collisions are pions. A large fraction (between 40 % and 80 %) of these pions arise from resonances. Since the resonances have finite lifetimes and momenta, their decay products are created in general outside the production region of the "direct" pions (i.e., pions produced directly from the source) and resonances. As a consequence, the two-particle correlation function of pions reflects not only the geometry of the (primary) source but also the momentum spectra and lifetimes of resonances. Kaons are much less affected by this circumstances; however, correlation experiments with kaons are much more difficult because of the low statistics.

Given the complexity of the problem, there are at present only two main methods to study the influence of resonances on BEC: Monte-Carlo calculations [15] and hydrodynamical calculations [16]. I shall mention in the following some results obtained via hydrodynamics not only because this method is more related to the subject of this meeting but also because hydrodynamics is the only way to obtain information about the equation of state.

The correlation function of two identical particles of momenta $p_1$ and $p_2$ can be written as
$$C_2(p_1, p_2) = 1 + \frac{A_{12}A_{21}}{A_{11}A_{22}}, \qquad (10)$$

where the matrix elements $A_{ij}$ are given in terms of source functions $g(x,k)$ as

$$A_{ij} = \sqrt{E_i E_j} < a^\dagger(p_i)a(p_j) > = \int d^4x g(x_\mu, k^\mu) e^{iq^\mu x_\mu}. \tag{11}$$

Here, $a^\dagger(p)$ and $a(p)$ are the creation operator and the annihilation operator of a particle of momentum $p$, and the four-momenta $k^\mu = \frac{1}{2}(p_i^\mu + p_j^\mu)$ and $q^\mu = p_i^\mu - p_j^\mu)$ are the average momentum and the relative momentum of the particle pair, respectively. The interpretation of the source function $g(x_\mu, p^\mu)$ as the quantum analogue of the mean number of particles of momentum $p^\mu$ at the space-time point $x_\mu$ enables us to decompose $g$ with respect to the origin of the produced hadrons. For instance, if the particles under consideration are two identical pions (e.g., two $\pi^-$), one has

$$g(x_\mu, p^\mu) = g_\pi^{dir}(x_\mu, p^\mu) + \sum_{res=\rho,\omega,\eta,...} g_{res\to\pi}(x_\mu, p^\mu), \tag{12}$$

where the labels *dir* and *res* $\to \pi$ refer to direct pions and to pions which are produced through the decay of resonances (such as $\rho, \omega, \eta$, etc.), respectively. The contribution from a resonance decay $g_{res\to\pi}(x_\mu, p^\mu)$ can be expressed in terms of the source function of that resonance itself, $g_{res}^{dir}(x_\mu^*, p^{*\mu})$, as follows (from here on, quantities related to a resonance will be labeled with a superscript star). Consider a resonance of width $\Gamma$, which is created at a space-time point $x_\mu^*$ and after a proper time $\tau$ decays into $\pi + X$ at $x^\mu = (x^{*\mu} + (\tau/m^*)p^{*\mu})$. As the influences of the decay time $\tau$ are described by the probability distribution $\Gamma exp(-\Gamma\tau)$, one obtains

$$g_{res\to\pi}(x_\mu, p\mu) \tag{13}$$
$$= \int \frac{d^3p^*}{E^*} \int d^4x^* \int_0^\infty d\tau \Gamma exp(-\Gamma\tau) \delta^4\left[x^\mu - \left[x^{*\mu} + \frac{\tau}{m^*}p^{*\mu}\right]\right] \Phi_{res\to\pi} g_{res}^{dir},$$

where $\Phi_{res\to\pi}(p^{*\mu}, p^\mu)$ describes the probability for a resonance of momentum $p^{*\mu}$ to produce a pion of momentum $p^\mu$.

Assuming that the decay is governed by phase space one can determine the functions $\Phi$. The source functions $g$ are then calculated via hydrodynamics assuming a freeze-out at a given temperature $T_f$. It is worthwhile mentioning that the hydrodynamical code used is a fully 3 dimensional one. The importance of 3 solution resides in the fact that the transverse flow affects drastically not so much the transverse dimension of the system but rather the longitudinal one [17].

Among the results obtained one should mention a significant difference in the effective longitudiinal radii extracted from $\pi^-\pi^-$ correlations and $K^-K^-$ correlations.

In the central region one has $R_{\|}^{K^-}/R_{\|}^{\pi^-} \simeq \frac{1}{2}$. This is due to the fact mentioned above that kaons are much less affected by resonances than pions. This effect has been experimentally confirmed by the Na44 collaboration. Another effect predicted by hydrodynamics and which has been apparently confirmed by the Na35 collaboration is the specific dependence of effective radii on rapidity.

### 2.2.4 New effects in Bose Einstein correlations; breakdown of the wave function approach

We come now to what is one of the most important developments in the field of BEC in the last years. It consists in the realization that the conventional understanding of BEC, namely that it is a correlation restricted to identical particles has to be qualified. In particular there exists a quantum statistical correlation also between particles and antiparticles. Although this new effect is quantitatively small and has not yet been observed experimentally, it is of far reaching significance. As a matter of fact, we are only at the beginning of the understanding of its implications. To judge the impact of this finding it is enough to mention that since the end of 1991 when this surprising effect was found [18], further three different derivations of the effect were given [19],[20],[9](cf. also ref. [21] which is based on ref.[9]). The initial motivation of the authors of these three papers was that the results of [18] seemed so surprising that they were "unbelievable" [19]. The surprising element is due to the fact that in most conventional approaches to BEC one uses an explicit symmetrization of the wave function which describes the two particle state and such a symmetrization applies of course only for identical particles. That the wave function approach is not always a convenient method for the study of BEC was known for many years because it describes only a state with a fixed number of particles, while experimentally the number of particles is usually not kept fixed (one measures inclusive probabilities rather than exclusive ones). An alternative to the wave function approach is the classical current approach, which is based on field theory i.e. second quantization. Strangely enough, alhough this alternative was known for many years, it had not been applied until 1991 but to identical particles, although, as will be shown below, it can be applied for particles-antiparticles as well.

Let us start with with some random currents $J_i(x)$ which create pions $\pi_i(x)$. The indices $i = 1, 2, 3$ refer to the isospin components. The simplest form of the Lagrangian for the field-current interaction reads

$$\mathcal{L}_{int}(x) = J_1(x)\pi_1(x) + J_2(x)\pi_2(x) + J_3(x)\pi_3(x)$$
$$= J_+(x)\pi^-(x) + J_-(x)\pi^+(x) + J_0(x)\pi^0(x) \tag{14}$$

Charged pions
$$\pi^\pm = \frac{1}{\sqrt{2}}(\pi_1 \pm i\pi_2) \tag{15}$$

are created by complex conjugated currents $J_\mp(x)$, and neutral pions $\pi^0 = \pi_3$ are created by real currents $J_0(x) = J_3(x)$.

We use now the general space-time correlator eq.(4) or eq.(5) and the space-time distribution of the source $f(x)$ and assume a Gaussian form of the density matrix.

In momentum space there are then two types of current correlators

$$\begin{aligned} <J_m^*(k_1)J_m(k_2)> &= <J_+^*(k_1)J_+(k_2)> = F(k_1, k_2) \\ <J_m(k_1)J_m(k_2)> &= F(-k_1, k_2) \qquad (m = 1, 2, 3) \end{aligned} \tag{16}$$

where $F$ has been defined in eq. (7).

Introducing now the creation and annihilation operators $a_m^\dagger$ and $a_m$ of the pion field one obtains

$$F(k_1, k_2) = (2\pi)^3 \sqrt{4E_1 E_2} \; <a_m^\dagger(k_1)a_m(k_2)> \tag{17}$$

$$F(-k_1, k_2) = -(2\pi)^3 \sqrt{4E_1 E_2} \; <a_m(k_1)a_m(k_2)> \tag{18}$$

While eq.(17) leads to the usual $\pi^-\pi^-$ or $\pi^+\pi^+$ correlation, Eq.(18) leads to the "new" $\pi^+\pi^-$ correlation. Related to this new correlation is the fact that $\pi^0$-$\pi^0$ correlations are different from charged pion correlations. Introducing the normalized current correlators

$$d_{rs} = \frac{F(k_r, k_s)}{[F(k_r, k_r) \cdot F(k_s, k_s)]^{\frac{1}{2}}}, \quad \tilde{d}_{rs} = \frac{F(k_r, -k_s)}{[F(k_r, k_r) \cdot F(k_s, k_s)]^{\frac{1}{2}}} \tag{19}$$

one has

$$\begin{aligned} C_2^{++}(\vec{k}_1, \vec{k}_2) &= 1 + |d_{12}|^2, \\ C_2^{+-}(\vec{k}_1, \vec{k}_2) &= 1 + |\tilde{d}_{12}|^2, \\ C_2^{+0}(\vec{k}_1, \vec{k}_2) &= 1, \\ C_2^{00}(\vec{k}_1, \vec{k}_2) &= 1 + |d_{12}|^2 + |\tilde{d}_{12}|^2 \end{aligned} \tag{20}$$

An estimate of the new terms shows that these are quite small and of the order of $\exp[-(E_1 + E_2)^2\tau^2]$ where $E^2 = k^2 + m^2$ and $\tau$ is the lifetime. Their importance however must not be underestimated and it consists, among other things in the fact that according to eq. (18) expectation values of the product of two creation operators do not vanish, which is a manifestation of squeezed states i.e. two particle coherent states, in analogy to usual coherent states for which the expectation value of a single annihilation operator is nonzero. It follows herefrom that squeezed states in particle physics are a quite natural phenomenon.This is at a first look very surprising, because in optics where these states were discovered for the first time, one has to prepare the system in a very special way in order to get them. At a closer look, however, one realizes that the actual observation of these effects in particle physics is not easy at all, because of the constraints imposed by the smallness of momenta of the particles. On the other hand, the fact that the new terms represent an *anti*correlation rather than the conventional corelation may help in their detection.

Another derivation of the new effects this time in the string model, is due to Bowler [19]. Although this derivation is only qualitative, it is interesting because it comes from a different point of view. Also the discussion of the relationship between the classical current approach and the string model is quite instructive.

Last but not least the work of Sinyukov and collaborators is worth mentioning. In ref.[20] it is argued that when using overlapping wave packets in the first quantization the appearance of the new terms is possible albeit in the view of the authors, improbable. On the other hand in second quantization by considering the density matrix appropriate for a system in local (but not global) equilibrium and in particular in boost invariant hydrodynamics a completely different picture emerges (cf.refs. [21], [9]). It turns out that terms of the form eq.(18) must exist and their form agrees with that derived in [18].[3]

An important corollary of [9] is the fact that the single inclusive distribution is also affected by the existence of squeezed states. This follows from the modification of the vacuum by the presence of terms of the form $<aa>$ and $<a^\dagger a^\dagger>$ and should lead to a strong enhancement of the low $p_\perp$ spectrum.

Finally one should mention a general quantum field theoretical derivation of the

---

[3]Ref. [21] contains an erroneous statement about the result of ref.[18] refering to the maximum of the correlation function for identical neutral pions.It is stated that in ref.[18] $C_{max}(\pi^0\pi^0) = 3$ "independent of particle mass and source size". This is obviously in contradiction with eqs.(18,22,23) of ref.[18]. From these eqs. it is clear that the maximum of the correlation function depends on mass, radius and lifetime and only in very special cases like zero lifetime or vanishing momenta and masses the current approach leads to $C_{max} = 3$.

particle-antiparticle correlation [22] which proves explicitely that this effect is not an artifact of the classical current formalism, is not restricted to dense media and is not specific to any space-time symmetry assumption as the results of ref.[9] might suggest. Furthermore ref.[22] provides an estimate of the magnitude of quantum corrections to the classsical current formalism

At the end of this detour into BEC one may ask what are the implications of the recent findings for the starting point of this talk, the Hagedorn mass spectrum. A tentative answer to this question is to say that the parameters of the spectrum, i.e the Hagedorn temperature and the power of the preexponential factor must reflect these effects. It is the task of a future theory of strong interactions to show this. We may have to wait until Hagedorn's 80-th anniversary to see this.

I am indebted to Michael Plümer for valuable discussions and reading the manuscript.